\documentclass[sigconf]{acmart}

\def\BibTeX{{\rm B\kern-.05em{\sc i\kern-.025em b}\kern-.08emT\kern-.1667em\lower.7ex\hbox{E}\kern-.125emX}}
    
\copyrightyear{2019}
\acmYear{2019}
\setcopyright{acmcopyright}
\acmConference[GLSVLSI '19]{Great Lakes Symposium on VLSI 2019}{May 9--11,
2019}{Tysons Corner, VA, USA}
\acmBooktitle{Great Lakes Symposium on VLSI 2019 (GLSVLSI '19), May 9--11, 2019,
Tysons Corner, VA, USA}
\acmPrice{15.00}
\acmDOI{10.1145/3299874.3319452}
\acmISBN{978-1-4503-6252-8/19/05}
 
\usepackage{multirow}
\usepackage{comment}
\usepackage{xcolor}
\usepackage{etoolbox}
\newbool{inccomment}
\setbool{inccomment}{true}

\newcommand{\hl}[1]{\ifbool{inccomment}{{\color{blue}#1}}{ }}

\begin{document}

%
\title[An Overview of eNVM-based PIM]{An Overview of In-memory Processing with Emerging Non-volatile Memory for Data-intensive Applications}

%
\author{Bing Li}
\authornote{This author is supported by NAS Associate Fellowship Award.}
\email{bing.li.ece@duke.edu}
\orcid{0000-0003-0732-2267}
\affiliation{%
  \institution{ECE Dept., Duke University, US}
  \institution{Army Research Office, Research Triangle Park, US}
}
\author{Bonan Yan}
\affiliation{%
  \institution{ECE Dept., Duke University\\Durham, North Carolina, US}
  }
\email{bonan.yan@duke.edu}

\author{Hai ``Helen'' Li}
\affiliation{%
  \institution{ECE Dept., Duke University\\Durham, North Carolina, US}
  }
\email{hai.li@duke.edu}

%
\renewcommand{\shortauthors}{Bing and Bonan, et al.}

%
\begin{abstract}
The conventional von Neumann architecture has been revealed as a major performance and energy bottleneck for rising data-intensive applications. 
The decade-old idea of leveraging in-memory processing to eliminate substantial data movements has returned and led extensive research activities. 
The effectiveness of in-memory processing  heavily relies on the memory scalability, which cannot be satisfied by traditional memory technologies. 
Emerging non-volatile memories (eNVMs) that pose appealing qualities such as excellent scaling and low energy consumption, on the other hand, have been heavily investigated and explored for realizing in-memory processing architecture. 
In this paper, we summarize the recent research progress in eNVM-based in-memory processing from various aspects, including the adopted memory technologies, locations of the in-memory processing in system, supported arithmetics, as well as applied applications. 
\end{abstract}


%
%
\maketitle

\section{Introduction}
In current era with data explosion, deep neural network (DNN) models are used to process various applications that explore a large amount of information in different data formats. 
Executing such data-intensive applications on conventional \textit{von Neumann} systems causes massive data movements between processors and memory elements and induces significant performance and energy overheads. 
After decades since it was proposed first time~\cite{patterson1997case,draper2002architecture}, the concept of in-memory processing returns and evokes many innovative solutions.  
Different from the conventional computing paradigm where data and computing are decoupled, in-memory processing architecture pulls data close to processing elements to reduce the amount of data movement and minimize the computation cost. 

\begin{table}[b]
\vspace{-6pt}
\caption{Emerging Non-volatile Memory Comparison}
\small
\label{tab:envm}
\begin{center}
\vspace{-12pt}
\begin{tabular}{cccccc}
\toprule
                & SRAM & DRAM & STT-RAM & PCM & ReRAM\\\midrule
Cell Size  &   \multirow{2}{*}{>100}   &  \multirow{2}{*}{6-10}    &    \multirow{2}{*}{6-50}     &  \multirow{2}{*}{4-30}   & \multirow{2}{*}{$\leq$2} \\
(F$^2$) & & & & & \\\midrule
Multibit &   1   &  1    &    1     &  >2   & 2-7 \\\midrule
Endurance &   >10$^{16}$   &  >10$^{16}$    &    >10$^{15}$     &  10$^8$-10$^{15}$   & 10$^8$-10$^{12}$ \\\midrule
Read Time  & \multirow{2}{*}{$\sim$1}& \multirow{2}{*}{$\sim$10} & \multirow{2}{*}{$<$10} & \multirow{2}{*}{$<$10} & \multirow{2}{*}{$<$10} \\
(ns)  & & & & & \\\midrule
Write Time & \multirow{2}{*}{$\sim$1} & \multirow{2}{*}{$\sim$10} & \multirow{2}{*}{$<$10} & \multirow{2}{*}{$~$50} & \multirow{2}{*}{$<$10} \\
(ns)  & & & & & \\\midrule
Write Energy      & \multirow{2}{*}{$\sim10^{-15}$} &  \multirow{2}{*}{$\sim10^{-14}$} & \multirow{2}{*}{$\sim10^{-13}$} &  \multirow{2}{*}{$\sim10^{-11}$} & \multirow{2}{*}{$\sim10^{-13}$}  \\
 (J/bit) & & & & & \\
\bottomrule
\end{tabular}

\end{center}

\footnotesize\emph{Source:}~\cite{Micron_power,burr2016recent,mittal2015survey,yu2016emerging}. \emph{Note:} F represents the feature size.
\end{table}

\begin{table*}[t]
\caption{An Overview of In-memory Processing Designs}
\small
\vspace{-12pt}
\label{tab:overview}
\begin{center}
\begin{tabular}{cc|c|c|ccc}
\toprule
      \multicolumn{2}{c|}{Works} & Types & Locations & Design Levels & Functions & Applications\\\midrule
Guo~\textit{et al.}~\cite{guo2010resistive} & 2010 &   \multirow{6}{*}{STT-RAM}   &  Cache    &    Circuit; System     &  Logic; Arithmetic &  Generic \\
AC-DIMM~\cite{guo2013ac} & 2013 &    &  Main Memory    &    Circuit; System     &  Associative &  Generic \\
Kang \textit{et al.}~\cite{kang2017memory} & 2017 &   &  -    &    Circuit     &  Logic &  Bitmap\\
STT-CiM~\cite{jain2018computing} & 2018 &     &  Scratchpad    &    Circuit; System     &  Logic; Addition; Vector &  Generic \\
HielM~\cite{parveen2018hielm} & 2018 &    &  -    &    Circuit; System  &  Logic & Encryption, Database\\
Pan \textit{et al.}~\cite{pan2018multilevel} & 2018 &     &  Co-processor    &    Circuit; System     &  Logic &  Binary CNN \\

\midrule
Cassinerio~\textit{et  al.}~\cite{cassinerio2013logic}& 2013 & \multirow{7}{*}{PCM}   &  -    &    Device  &  Logic &  - \\
Wright \textit{et al.}~\cite{wright2013beyond,wright2011arithmetic}& 2011, 2013 &    &  -    &    Device  &  Arithmetic &  - \\
Hosseini \textit{et al.}~\cite{hosseini2015accumulation}& 2015 &    &  -    &    Device  &  Arithmetic &  - \\
Pinatubo~\cite{li2016pinatubo} & 2015 &   & Main Memory    &  Circuit; System  &  Logic & Generic \\
Burr \textit{et al.}~\cite{burr2015experimental,burr2015large}& 2015 &   &  -    &    Circuit  &  MVM & DNN \\
Sebastian \textit{et al.}~\cite{sebastian2017temporal}& 2017 &   &  -    &    Circuit  &  MVM &  Unsupervised Learning \\
Le \textit{et al.}~\cite{le2017compressed,le2018mixed}& 2017, 2018 &   &  -    &    Circuit  &  MVM & Transfer Learning \\\midrule
MAGIC~\cite{kvatinsky2014magic}& 2014 & \multirow{6}{*}{ReRAM} & Co-processor & Circuit & Logic; Arithmetic & Adder \\ 
Bojnordi~\textit{et al.}\cite{bojnordi2016memristive}& 2016 & & Co-processor & System & MVM & Boltzmann machine \\
ISAAC~\cite{shafiee2016isaac} & 2016 & & Co-processor & System & MVM & CNN \\
PipeLayer~\cite{song2017pipelayer}& 2017 & & Co-processor & System & MVM & CNN \\
AtomLayer~\cite{qiao2018atomlayer}& 2018 & & Co-processor & System & MVM & CNN \\
GraphR~\cite{song2018graphr}& 2018 & & Co-processor & System & MVM & Graph \\

\bottomrule
\end{tabular}
\end{center}
\footnotesize \emph{Note}: MVM -- Matrix-Vector Multiplication; DNN -- Deep Neural Network.
\end{table*}

Benefiting from the recent advances in processing integration and memory technologies, many in-memory processing architectures have been developed.
These attempts can be cataloged into three groups:
(1) processing close to memory, aka near-data processing (NDP); 
(2) processing in traditional memory; and
(3) processing in emerging non-volatile memory (eNVM).

In enabling NDP, three-dimension (3D) integration 
is a key technology. 
For example, 3D DRAM is constituted by vertically stacking a set of DRAM dies on top of a CMOS logic die over through-silicon vias (TSVs).
There are a number of 3D DRAM-based NDP platforms that exploit the logic die to perform simple but common operations in data-intensive applications~\cite{pugsley2014ndc,farmahini2015nda,zhang2014top,gao2016hrl,gao2017tetris,kim2016neurocube,hosseini2015accumulation}. 
TSVs substantially shorten the distance between the logic and memory dies, increasing the data bandwidth and improving overall performance. 

Processing in memory directly performs computations in memory arrays so as to reduce data movement to a large extent. 
Prior works exploit traditional memory technologies such as DRAM and SRAM to complete the frequent operations appearing in data-intensive applications ~\cite{li2017drisa,deng2018dracc,zhang2017memory,agrawal2018x,kang2018multi}. 
However, as the scaling of traditional memory technologies is approaching the physical limit, it is difficult to provide sufficient computing and storage capacity for data-intensive applications. 
Moreover, big cell size and high leakage power of traditional memory lead to large design area and energy consumption~\cite{patel2017reach,park2012future,mittal2014survey}.

In recent years, eNVMs that demonstrate excellent scaling and near-zero leakage power are emerged as promising candidates for future trend. 
Table~\ref{tab:envm} compares traditional DRAM and SRAM with a few popular eNVM technologies, including spin-transfer torque RAM (STT-RAM), phase-change memory (PCM) and resistive RAM (ReRAM). 
Among these eNVM technologies, STT-RAM shows the fastest access speed and the lowest energy consumption while the cell area is relatively larger~\cite{park2012future,sayed2018cross}.
Both PCMs~\cite{nirschl2007write,li2017power} and ReRAMs~\cite{wong2012metal,xu2015overcoming,li2019build} can store multiple logic bits in a single memory cell, demonstrating superior density with technology scaling.
In addition, they inherently support parallel data processing, which is uniquely beneficial to aforementioned data-intensive applications like DNNs. 
For the reason, extensive research efforts have been devoted to building in-memory processing using eNVM.

In this paper, we survey the recent progress in developing in-memory processing by leveraging the three mainstream eNVM technologies (STT-RAM, PCM and ReRAM). 
We present and discuss the difference and similarity of these works in terms of the supported functions, the location in architecture, the targeted applications, \textit{etc}.   
\section{Design Overview}
\begin{figure}[b]
\centering
\vspace{-6pt}
\includegraphics[width=\columnwidth]{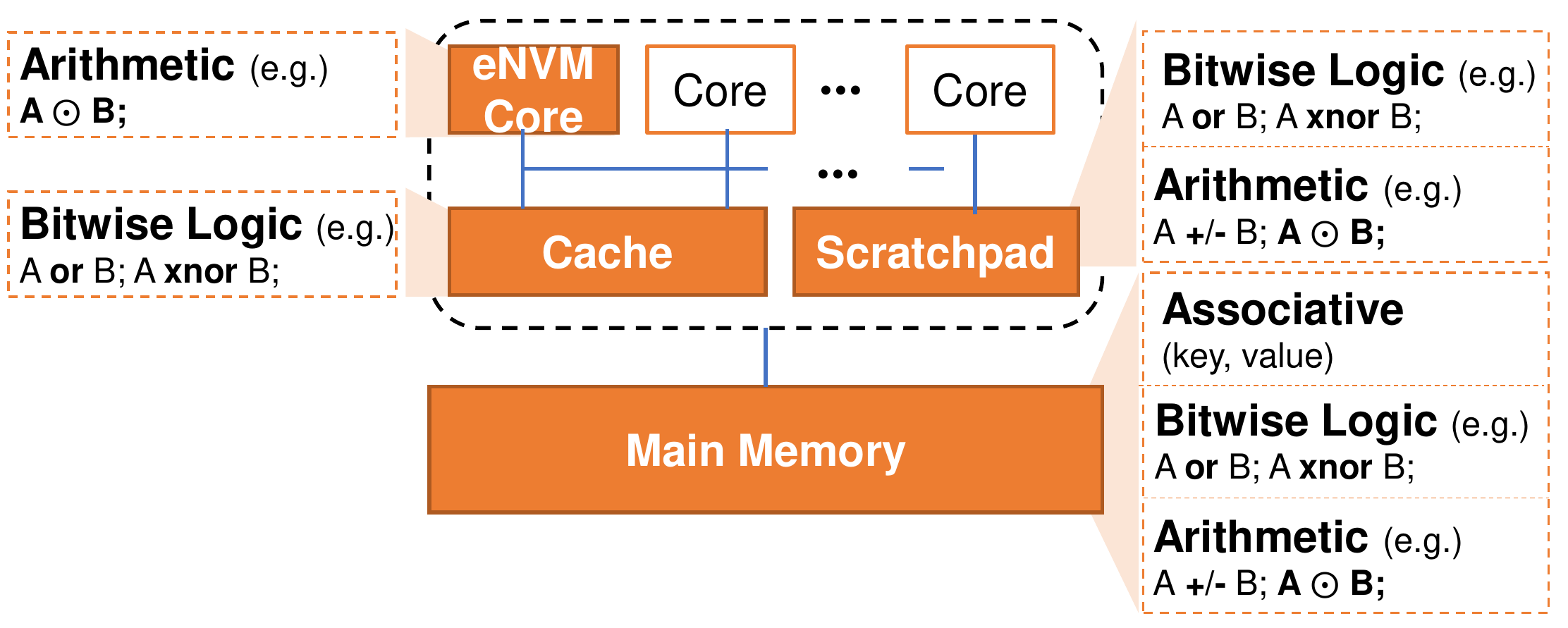}	
\vspace{-12pt}
\caption{Processing locations and function types.} 
\label{fig:overview} 
\end{figure}
Table~\ref{tab:overview} presents a summary of the latest eNVM-based in-memory processing designs reviewed in this paper. 
Each of them is classified according to the following five main categories. 
\begin{itemize}
    \item \emph{Type -} The types of memory technologies adopted in these works: \textbf{STT-RAM, PCM} or \textbf{ReRAM}. Since the features of memories are different with each other, the selection of memory type determines the types of computation to some extent.
    \item \emph{Location -} Where is the memory located in the computing architecture: \textbf{cache, main memory} or \textbf{scratchpad}? 
    eNVMs can be used as storage and/or computing unit. 
    Some works treat eNVM only as \textbf{co-processor} while some designate its location in memory hierarchy too. 
    \item \emph{Design Level -} The techniques in these works are carried out at different levels, such as \textbf{device, circuit} or \textbf{system}. 
    Some works proposed the novel writing method to perform calculations in memory cells~\cite{cassinerio2013logic,wright2013beyond,wright2011arithmetic,hosseini2015accumulation}. 
    Some techniques are achieved through the modifications of the readout or write circuits associated with memory arrays. 
    The system-level techniques would provide the interface and connection between the memory array and operation system so that the processing can be manipulated by applications.
    \item \emph{Functions -} The function types in these works can be divided into the following groups: \textbf{logic, arithmetic, associative, vector} and \textbf{matrix-vector multiplication (MVM)}. 
    A type of operation can be realized by different eNVMs, but the implementation details could vary significantly.
    \item \emph{Applications -} Most works provide advanced functions to support most data-intensive applications, which are grouped into~\textbf{generic}. 
    Some works carry out only the core operation in a number of applications such as~\textbf{encryption, database}, and \textbf{CNN}. 
    A few works complete simple and basic operations that are not designated to any targeted applications. 
\end{itemize}

\begin{figure*}[t]
\centering
\includegraphics[width=.75\textwidth]{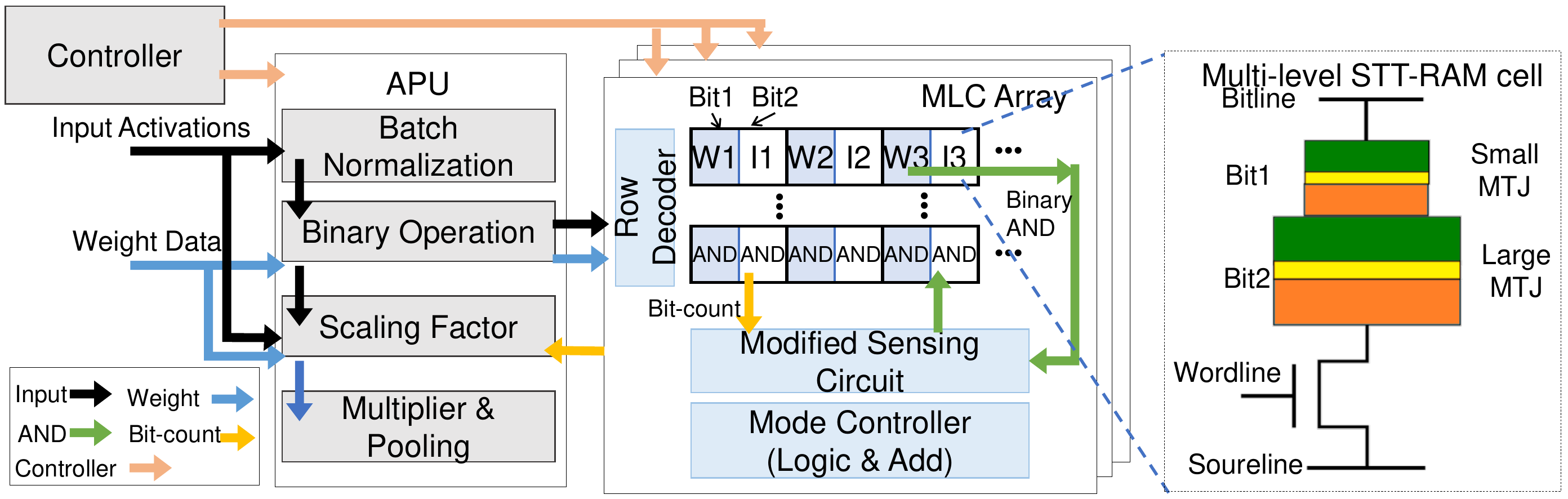}	
\vspace{-9pt}
\caption{The execution flow, computing array, and multilevel STT-RAM cell for convolutional layers of Binary CNN in~\cite{pan2018multilevel}.} 
\label{fig:sttbcnn} 
\vspace{-12pt}
\end{figure*}

Figure~\ref{fig:overview} depicts a high-level view of the classifications based on the location of the in-memory processing in system (eNVM core, cache, scratchpad memory, or main memory) and the type of supported functions (bitwise logic, arithmetic, or associative).

\section{eNVM-based In-memory Processing}
\subsection{Spin-Torque Transfer RAM (STT-RAM)}

STT-RAM consists of a magnetic tunnel junction (MTJ) device, which presents two resistance states depending on the relative magnetization orientation of the fixed and free ferromagnetic layers. 
Compared to other resistive memory devices, STT-RAM has faster write speed, lower write energy, and higher write endurance (refer Table~\ref{tab:envm}). 
Due to the limited resistance difference between these distinct resistance states of MTJ, it is hard to implement multi-bit storage in STT-RAM cells. 
So most of STT-RAM-based in-memory processing designs focus on the bit-wise operations.

\subsubsection{Associative and combinational logic}
Early works exploit the high density of STT-RAM to complete the associative computing and combinational logic~\cite{guo2010resistive,guo2013ac}.
These works achieve reduced cost relative to traditional memory technologies. 
For instance, \textbf{Guo~\textit{et al.}}~\cite{guo2010resistive} employ STT-RAM to construct \emph{look-up table} (LUT) and further realize the computing by cascading multiple LUTs. 
As such, the floating-point units are replaced by STT-RAM-based LUTs. 
The work successfully demonstrates the improved power and performance brought by STT-RAM technology, compared to multi-core CPU platform. 

\subsubsection{Bitwise logic operations}
Recent works~\cite{jaiswal2018situ,kang2017memory,jain2018computing,parveen2018hielm} explore the use of STT-RAM in accomplishing bitwise logic operations. 
Based on the basic logic function realization  by STT-RAM, advanced operations are implemented. 
Kang~\textit{et  al.}~\cite{kang2017memory}, STT-CiM~\cite{jain2018computing} and HieIM~\cite{parveen2018hielm} are taken as examples and introduced here.

\textbf{Kang~\textit{et  al.}}~\cite{kang2017memory} propose a STT-RAM chip which can process bitwise logic and store information. The operands reside in different rows of the same array. By simultaneously activating multiple rows, the bitwise operations are enabled and results are obtained through the modified readout periphery circuits, \textit{i.e.} sense amplifier (SA). The functionality of one logic operation can be controlled by modifying the bit in one control row. The chip can benefit some particular applications which involve intensive bitwise logic operations such as ~\emph{bitmap}. 
This work focuses on circuit design and functional evaluation.

\textbf{STT-CiM}~\cite{jain2018computing} extends the supported functions from bitwise logic to basic arithmetic and vector operations. 
At circuit level, the row decoders and SAs are enhanced to enable logic functions. 
Additional logic gates are integrated into the sense circuits to realize arithmetic operations. 
Two row decoders are used to active multiple rows where operands are located. 
The connected multiplex circuits are controlled by select signals to determine the desired operation type. 
When processing vector function, the vector outputs from STT-RAM arrays will be fed into reduction units and switched to the scalar value. 
At array level, authors analyze the impact of process variation on the computing results and deploy error correction scheme to enhance the reliability. 
At architectural and system levels, this work extends the instruction set to convey the operation command from applications to memory array. 
Through the extensions across multiple levels together, STT-CiM can be placed next to processor as on-chip scratchpad memory and applied for various data-intensive applications such as text processing, data compression, and digits recognition.

\textbf{HieIM}~\cite{parveen2018hielm} implements bulk bitwise operations in STT-RAM array too. 
Different from the above two designs, HieIM is more flexible and allows the computing to operate between any cells within the same array. Moreover, a data encryption engine based on HieIM is demonstrated, which consumes 51.5\% lower energy than the CMOS-based ASIC counterpart.

\subsubsection{Neural networks}
Thanks to the evolution of DNN models, convolution in binary convolutional neural network (BCNN) can be replaced with bitwise operations such as XNOR and bit-count~\cite{rastegari2016xnor}. \textbf{Pan~\textit{et al.}}~\cite{pan2018multilevel} build an accelerator based on multilevel STT-RAM (\textit{i.e.} two-bit cell) for BCNN. 
STT-RAM arrays are programmable and can be switched between memory mode and bitwise operation mode. 
Thereby, multi-functional STT-RAM arrays are exploited to process convolutional layers. 
In this design, the two bits of one cell associates with inputs and weights, respectively so that the logic and add operations are carried out within one cell. 
This integrates the compuational STT-RAM array with an auxiliary processing unit (APU) which processes other computational layers in CNNs and implements the BCNN accelerator. 
Figure~\ref{fig:sttbcnn} illustrates the computing array and execution flows. 
Compared to other eNVM-based counterparts, the STT-RAM based accelerator achieves significant performance and energy improvement. 

\begin{figure*}[t]
\centering
\includegraphics[width=.8\textwidth]{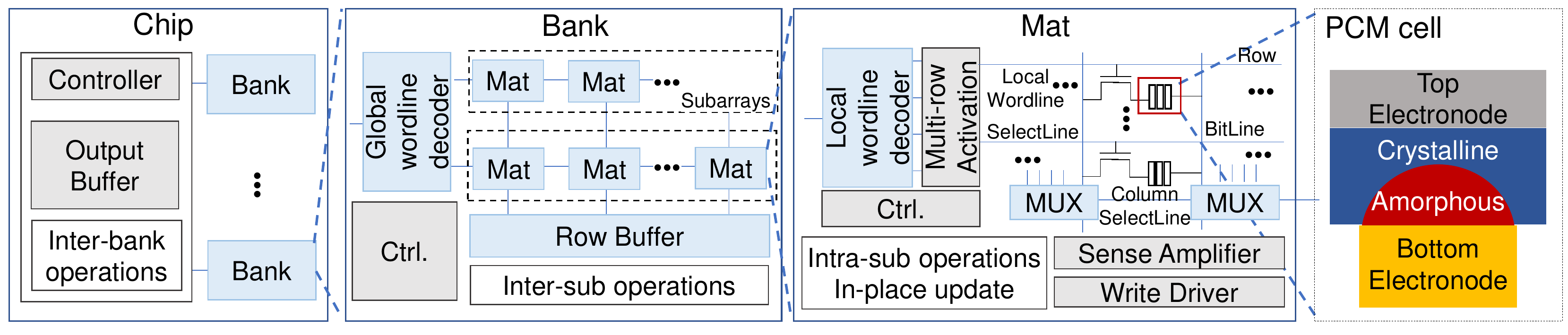}	
\vspace{-9pt}
\caption{Pinatubo~\cite{li2016pinatubo} and PCM cell architecture.} 
\vspace{-12pt}
\label{fig:pcm_pinatubo} 
\end{figure*}

\subsection{Phase Change Memory (PCM)}
PCM can store more than one bit of data per cell, by diving the overall resistance range into a few levels. 
What's more, the cell conductance exhibits a linear increase along with the number of programming (more exactly, SET) pulses~\cite{sebastian2018brain}. 
These key attributes of PCM are exploited to implement more complicate computations, such as the training of neural networks.
 
\subsubsection{Logic and basic arithmetic operations}
Phase change material manifests many physical attributes under various pulse amplitudes or duration, which have been exploited to realize computation.
For example, \textbf{Cassinerio~\textit{et  al.}}~\cite{cassinerio2013logic} leverage the resistance transition of phase change material and propose an initialize-compute-confirm scheme to implement Boolean logic operations within a single PCM cell. 
\textbf{Wright \textit{et al.}}~\cite{wright2013beyond,wright2011arithmetic} and \textbf{Hosseini \textit{et al.}}~\cite{hosseini2015accumulation} exploit the accumulative behavior of PCM material during programming and build an accumulator for arithmetic computations such as addition, subtraction, and parallel factoring. 
In these works, the partial and final results can be stored where computations are carried out. 
One single operation takes multiple cycles to complete as input operands sequentially enter. 
PCM cells are used to substitue for logic gates without revealing specific applications of interest.

\textbf{Sebastian \textit{et al.}}~\cite{sebastian2017temporal} exploit the physical dynamics of PCM material and propose computational PCM to perform the temporal correlation detection between stochastic binary processes. 
One process is encoded into a SET pulse whose amplitude or duration is proportional to the instantaneous sum of all processes and enters the assigned PCM device. 
By comparing the conductance of each device, the correlated processes can be identified.

\subsubsection{Matrix-vector multiplications \& machine learning}
Arranged in the crossbar structure, PCM devices can process analog matrix-vector multiplications, which have been intensively investigated~\cite{burr2015experimental,burr2015large,burr2016recent,le2017compressed,le2018mixed,ambrogio2018equivalent}. 
An element in an matrix can be corresponded to the conductance of a PCM device. 
With an PCM crossbar representing a matrix, the vector is encoded into the amplitudes or duration of voltage pulses applied along rows. Then, the currents along columns will be proportional to the results.
The positive and negative elements of the matrix could be stored in a pair of PCM devices.
When applying input signals to columns, the currents along rows denote the results of the vector multiplying with the transposed matrix. 
A 3-layer perceptron using PCMs trained with backpropagation on the MNIST database of handwritten digits can achieve the comparable accuracy with the software model~\cite{burr2015experimental,burr2015large}. 
Moreover, leveraging PCM-based in-memory processing for other complicated tasks are demonstrated, such as compressed sensing recovery~\cite{le2018mixed} and transfer learning~\cite{ambrogio2018equivalent}.



\subsubsection{System-level bitwise operations}
\textbf{Pinatubo}~\cite{li2016pinatubo} proposes a mechanism to perform bulk bitwise operations in PCM main memory. 
Read circuit and write driver is modified for Pinatubo processing logic functions. 
The operands are all stored in different rows in memory arrays. 
According to the locations that operands reside, Pinatubo has three computation modes: intra-subarray, inter-subarray and inter-bank (Figure~\ref{fig:pcm_pinatubo}). 
The rows associated with operands will be activated simultaneously when computing. 
Sense amplifiers are enhanced with more reference circuits to obtain the logic outputs which will be sent to I/O bus or another memory row. 
To bridge operating system and logic operations inside PCM, Pinatubo develops the programming model and run-time supports to ensure that operands are allocated to different memory rows. 
The design achieves 1.12$\times$ overall speedup, 1.11$\times$ overall energy saving over the conventional CPU. 

\begin{figure*}[t]
\centering
\includegraphics[width=.7\textwidth]{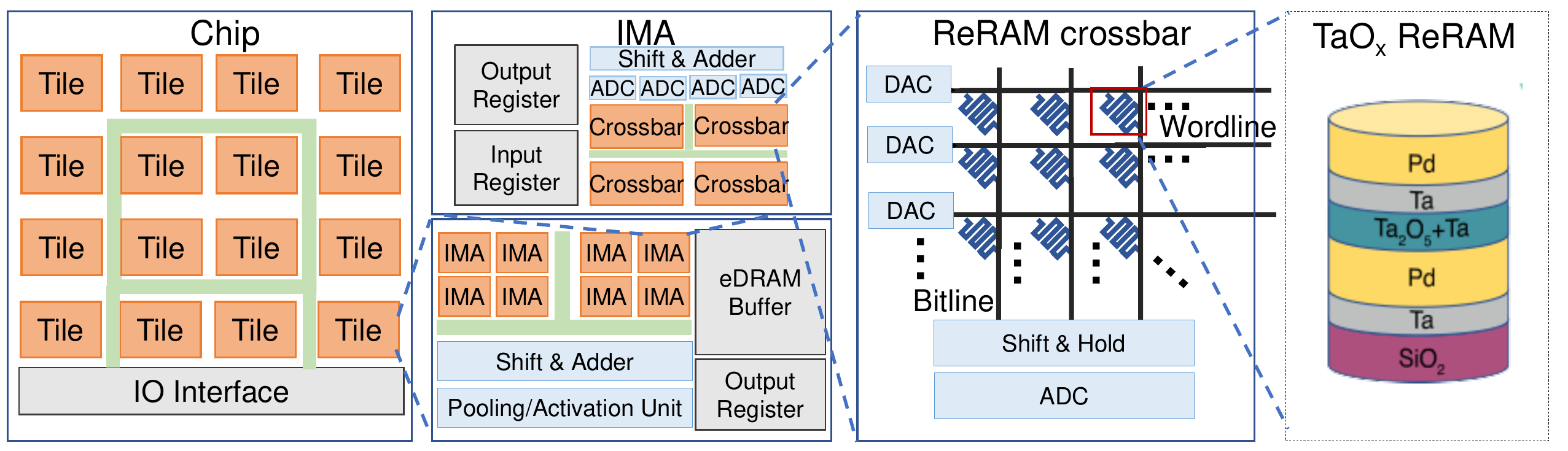}	
\vspace{-12pt}
\caption{ISAAC~\cite{shafiee2016isaac}, ReRAM crossbar, and ReRAM cell architecture.} 
\vspace{-12pt}
\label{fig:reram_shafiee} 
\end{figure*}

\subsection{Resistive RAM (ReRAM)}

The attractive features of high resistances and multi-level cell storage make ReRAM stand out from other emerging memory technologies to construct dense and low-power computing systems~\cite{yan2017understanding}.
A variety of ReRAM based computing systems have been proposed to demonstrate superior performances in different applications central to memory access reduction.

\subsubsection{Logic arithmetic operations} 
\textbf{MAGIC}~\cite{kvatinsky2014magic} proposes mechanisms to perform bitwise operations with the aid of binary ReRAM.
These schemes enable the integration of fundamental logic gates as well as complex arithmetic units, \textit{e.g.} multi-bit full adders, within a ReRAM array.

\subsubsection{Matrix-vector multiplications}

\textbf{ISAAC}~\cite{shafiee2016isaac} is an neural network accelerator based on ReRAM dot-product engine.
In ISAAC, ReRAM crossbar arrays both store weights in DNN and perform MVM with analog current and analog/digital converters (\textit{i.e.} DAC \& ADC).
Figure~\ref{fig:reram_shafiee} shows the top-down view of ISAAC chip.
A group of tiles are connected through on-chip network. Every tile is composed of eDARM buffers , several in-situ multiply-accumulators (IMA), output registers and the shift-adders. 
Pooling and activation units in tiles dedicate to the pooling and activation operations in neural network. 
Each IMA consists of a number of ReRAM crossbars, ADCs, the input/output registers, and shift-adders.
ISAAC exploits this integration of storage and computation for saving data (especially weights in filters) movement. 
The deeply pipelined flow of ISAAC focuses on the optimization in neural network inference.

A number of ReRAM-based CNN accelerators are proposed for boosting the system performance in training~\cite{song2017pipelayer,qiao2018atomlayer,cheng2017time}.
For example, \textbf{PipeLayer}~\cite{song2017pipelayer} balances the parallelism and throughput in training and inference based on both parallelism granularity and weight duplication. 
By eliminating the potential stalls as in ISAAC, PipeLayer yields an averagely 42.45$\times$ speedup and saves computation energy by 7.17$\times$ on average, compared to the massively parallel computing GPU platform.
\textbf{AtomLayer}~\cite{qiao2018atomlayer} attempts to provide a universal solution to enhance the efficiency during both training and inference. 
In this scheme, one network layer is executed at a time, \textit{i.e.} atomic layer, to solve the issues brought by the highly pipelined operations, such as pipeline bubbles, long single-layer latency, and high cost of data buffers. 
AtomLayer revises the mapping scheme of weights to ReRAM arrays and data reuse and further reduces the on-chip data buffer access aside from the reduction of memory accesses.
AtomLayer achieves 1.1$\times$ higher power efficiency than ISAAC in inference and 1.6$\times$ higher than PipeLayer in training, and its footprint shrinks 15$\times$ averagely with the reduction of on-chip buffers.

ReRAM array's parallel computing nature is capable of building accelerators for special computing models other than neural networks. 
\textbf{GraphR}~\cite{song2018graphr} is a ReRAM-based graph processing accelerator to solve the poor locality and high-bandwidth requirement in graph processing.
The ReRAM crossbar based graph engines offers low-cost hardware implementation to realize power-efficient graph processing acceleration.

Beyond convolutional computing engine, a number of works utilize ReRAM crossbars to support different computations and applications ~\cite{bojnordi2016memristive,li2018reram,yin2018parallel,fan2019red}.
For instance, \textbf{Bojnordi \textit{et al.}}~\cite{bojnordi2016memristive} implement the restricted Boltzmann machine with ReRAM arrays.
The model of Boltzmann machine has been used to train deep neural networks with vast training samples.
With the help of current summation circuit and reduction unit, large networks are reshaped to fit into ReRAM arrays, where in-situ computing operations are executed.
Compared with conventional multi-core systems, ReRAM-based Boltzmann machine achieves 57$\times$ higher performance and 25$\times$ lower energy consumption without degrading the quality of solutions to optimization problems.

\section{Conclusion}

In this work, we gave an overview of recent works on eNVM-based in-memory processing that
minimizes the cost of memory access and is expected to be meet the requirements of data-intensive applications.
Emerging non-volatile memories (eNVMs) have advantages of low-power, high-density, superior scaling and inherent computing capability. Hence, numerous research works have been carried out to develop eNVM-based in-memory processing architectures. 
We summarize and discuss the types of eNVMs that have been adopted in in-memory processing designs, as well as a variety of implemented functions and supported applications. 
Because each type of eNVMs has distinct strengths and weaknesses, the selection of eNVM technology shall consider the specific requirements of applications. 
Following the progress of material science and device processing techniques, we anticipate continuous improvement in reliability, read/write speed, and energy efficiency of eNVM technologies. 
We believe the collaborative researches across various levels including device, circuit, system and applications, are essential to move eNVM-based in-memory processing towards commercial production.

%
\begin{acks}
Bing Li acknowledges the National Academy of Sciences (NAS), USA for awarding the NRC research fellowship. Any opinions, findings and conclusions or recommendations expressed in this material are those of the authors and do not necessarily reflect the views of NAS or their contractors.
\end{acks}

%
\bibliographystyle{unsrt}
\bibliography{0_main}

\end{document}